\documentclass[10pt]{revtex4}
\usepackage{amsmath,amssymb}
\usepackage[pdftex]{graphicx}
\usepackage{subfigure}
\usepackage{bm}
\usepackage{epstopdf}
\begin{document}
\title{Energetic cooling below the BEC transition: a quantum
kinetic description within the Bogoliubov approximation}

\author{A. Camacho-Guardian, M. Mendoza-L\'opez, V. Romero-Roch\'{\i}n and R. Paredes\footnote{Corresponding author: rosario@fisica.unam.mx}}                   

\affiliation{ 
 Instituto de F\'{\i}sica, Universidad
Nacional Aut\'onoma de M\'exico, Apartado Postal 20-364, M\'exico D.
F. 01000, Mexico. }


\date{\today}
\begin{abstract}

The dynamics of Bose-Einstein condensation in a three-dimensional harmonic trap is studied 
explicitly including the Bogoliubov approximation for temperatures below the critical one.
To model the evolution towards equilibrium at each cooling step, we derive quantum kinetic equations that describe the dynamics of the gas for temperatures above and below the transition temperature. These equations, valid in the Born and Markov approximations, consider the essential role of the chemical potential as the main parameter that signals the transition. The kinetic equation that describes the growth of the condensate below the transition temperature is derived within the Bogoliubov approximation. To illustrate our results we propose an energetic cooling protocol and simulate the whole sequence of the formation of a condensate.
\end{abstract}

\date{\today}
\maketitle

\section{Introduction}
\label{intro}
The observation of Bose-Einstein condensation (BEC) in 1995 was the culmination of a series of experiments carried out with atomic neutral gases \cite{Cornell,Ketterle} in which macroscopic samples of matter were cool down until quantum degeneracy was reached.
As it is well known, the method of evaporative cooling was the essential technique to reach temperatures below the $\mu$K degrees, necessary for the realization of BEC. This technique works in several stages by lowering the temperature at each step as highly energetic particles are removed from the system. Condensation in macroscopic samples was finally attained at 150 nK at a density of 2.5 $\times$ $10^{12}$ cm$^{-3}$ for $^{87}$Rb atoms \cite{Cornell}, and below $2$ $\mu$K at  a density of 1 $\times$ $10^{14}$ cm$^{-3}$ for $^{23}$Na atoms \cite{Ketterle}.

While there is well established knowledge of ground state and non-equilibrium properties of  trapped BEC, mostly within the Hartree-Fock-Bogoliubov mean-field approach \cite{Stringari,Pethick}, those predictions are restricted to the very low temperature regime where the transition to BEC has already occurred. Thus, there is interest in describing the dynamics or kinetics of the formation of the condensate, taking into account the essential mechanism of atomic collisions that end up equilibrating the system once, say, a evaporative cooling step has been performed.

The incorporation of collisional effect has been the main goal in all the formulated kinetic theories so far  that describe the dynamics during the formation of a Bose-Einstein condensate \cite{Holland,Walser1,Walser2,Gardiner,Yamashita, Drummond,Morandi}. The present work is within such a  dynamical description context and it is particularly concerned with the derivation, in the weak coupling limit and the Markov approximations, of irreversible master equations for the one body reduced density matrix operator. To be precise, our aim is the description of the time evolution of the occupation numbers as the BEC transition takes place.

Similarly to earlier kinetic formulations, \cite{Holland,Walser1,Walser2,Gardiner} we work in the Born and Markov approximations to straightforwardly derive irreversible master equations that describe the dynamics during the formation of a condensate. As we explain below, we have added the important contribution of Bogoliubov to take into account the role played by the condensate below the transition temperature. We derive kinetic equations appropriate for temperatures $T > T_c$ and for $T < T_c$. The equations describing the relaxation towards equilibrium for $T>T_c$ allow us to demonstrate that the ideal Bose-Einstein distribution with negative chemical potential is the stationary solution for both the ground and excited states, and that the total number of particles as well as the total energy are conserved along the thermalization of each cooling step. However, since this  equation is unable of describing the dynamics for the ground and excited states when the chemical potential becomes zero, that is, for the dynamics below $T_c$, we derive a novel kinetic equation to properly describe the evolution in time of a  weakly interacting Bose gas for $T<T_c$. In particular, we consider the fact that when the chemical potential becomes zero, the particles in the separate ground and excited states are not conserved. Thus, we propose to replace the interaction term of the master equation, accounting for every single collision event, by an interaction potential in which the colliding particles arising from the ground state are treated within the Bogoliubov approximation \cite{Bijlisma,Fetter}. Technically speaking, we shall consider that the creation and annihilation operators involving the ground state $a_0^\dagger$ and $a_0$ can be substituted by $\sqrt{N_0}$, with $N_0$ the number of particles in the condensate. Such a practice is well justified by the fact that the number of of particles in the ground state, when the chemical potential is zero, becomes macroscopic, i.e. $N_0 \sim N$ and thus the matrix elements of the Hamiltonian in the number-occupation states appear simply as $\sqrt{N_0}$. This procedure leads us to obtain a new kinetic equation for the particles in the excited states, exhibiting the correct limits. The equilibration after each cooling stage can thus be analyzed for all temperatures, above and below $T_c$.

For the numerical simulation of the kinetic equations we propose here a protocol of energetic cooling. This scheme consists of several cooling steps in which the temperature in the closed system is decreased while maintaining constant the number of particles. The temperature at each cooling step is lowered by transferring particles from higher energy levels and distributing them in levels with lower energy. Interatomic collisions re-thermalize the gas to a lower temperature. In contrast with the evaporative cooling technique, this scheme prevents the system against decreasing the number of particles, yet, by reducing its energy, its density at the center of the trap increases and, thus, degeneracy can be reached. This protocol may represent an alternative route for the experimentalists to achieve the degeneration regime with a much larger number of atoms than usual. Although it is not the purpose of this article, we suggest that the transferring to lower energy levels may be implemented by means of impulsive anti-Stokes stimulated Raman scattering \cite{Boyd} with properly tailored ultrashort microwave pulses, similarly to techniques used with light in single-molecule wavepacket interferometry \cite{Scherer,Cola}.

This work is organized as follows. In section II we present the derivation of the kinetic master equations describing the growth dynamics of the condensate for temperatures above and below the critical temperature defining the transition. In section III we present numerical results that model the energetic cooling process as several stages are performed in a bosonic gas towards the formation of a condensate. Finally, in section IV a summary of this work is given.

\section{Quantum Boltzmann equations for the formation of a condensate}
\label{section1}

As envisaged by Einstein and stated in textbooks, there are two routes to achieve the condensation in a macroscopic system. One is by lowering the temperature and maintaining constant the density and the other is by increasing the density and keeping constant the temperature. The implementation of the first method requires to put the closed system in thermal contact with a reservoir at a lower temperature, wait for the system to reach thermal equilibrium and then repeat the scheme until the system reaches the transition temperature. The second route conceives an open system that increases its density being in contact with a reservoir at fixed temperature. As before, the system reaches an equilibrium state after a relaxation time, and again is placed in contact with a reservoir with larger chemical potential, as part of the sequence, until the transition is reached. It is important to note that both procedures ensure that the onset of condensation may be achieved in a reversible way. In practice none of these methods is performed in the laboratory. 

In actual experiments, the observation of the BEC transition requires to perform several cooling stages in a thermally isolated Bose gas, in which a given equilibrium state is perturbed reducing its energy (and the number of atoms in evaporative cooling) and then, by a necessarily irreversible process, the system reaches a new equilibrium state with a lower temperature. This cooling is repeated until the system attains the critical temperature $T_c$ at which the transition to condensation occurs. Our aim in this work is to describe, through quantum kinetic equations, the dynamical process that takes place during the thermalization at each cooling step. To accomplish this goal in this section we derive the kinetic equations that describe the relaxation process towards the stationary or equilibrium state in each cooling step. As we shall see, such a description requires to take into account that the cooling process must be divided in two parts, one for temperatures above the critical temperature $T_c$  at which the gas is in the normal phase, and the other for temperatures below $T_c$ when the BEC transition to occurs.

\subsection{Kinetic theory for temperatures above $T_c$}
\label{sectionA}

The Hamiltonian that describes the system consists of an ideal term $H_0$ that takes into account a isotropic harmonic confinement of the atoms, perturbed by a weakly interacting part $V$ that models the collisions among the particles. In a second quantization formalism these terms are: 

\begin{eqnarray}
&&H= \sum _{\vec{i}}  \epsilon_{\vec{i}} a^{\dag}_{\vec{i}}a_{\vec{i}}+\frac{1}{2} \sum_{\vec{i} \vec{j}\vec{k}\vec{l}} \langle \vec{i} \vec{j} \vert V \vert \vec{k} \vec{l} \rangle a^{\dag}_{\vec{i}} a^{\dag}_{\vec{j}}a_{\vec{k}}a_{\vec{l}}=H_0+V
\label{ideal}
\end{eqnarray}
where $\epsilon_{\vec{i}}$ are the eigenenergies of a single particle confined in a harmonic potential of frequency $\omega$ in three dimensions, $\epsilon_{\vec{i}}=\hbar \omega \left( i_x + i_y + i_z  \right )$. Since the energy of each single particle is characterized by three quantum numbers $i_x$, $i_y$, and $i_z$ their corresponding quantum state is denoted as a vector $\vec{i}$. $a ^{\dag}_{\vec{i}}$ and $a_{\vec{i}}$ are the operators that create and destroy particles in the state $\vec{i}$. In the expression for single particle energy $\epsilon_{\vec{i}}$ we have omitted the ground state energy $\epsilon_0 =\frac{3}{2} \hbar \omega$, and as we recall below, this fixes the BEC transition at $\mu=0$.  
Writing $i =i_x + i_y +i_z$ the degeneracy of each energy is given by
\begin{equation*}\label{degeneracy}
g_i = \frac{1}{2} \left( i +1 \right) \left( i + 2 \right).
\end{equation*}  

Regarding the interacting term, we shall assume that the ultracold gas is very diluted and consequently only $s$-wave binary collisions are relevant for the dynamics\cite{Weiner,Pethick,Ketterle2}. This two-body interaction can be modeled by a contact potential whose interaction strength is proportional to the $s$-wave scattering length $a_s$, assumed positive throughout,

\begin{equation}
V(\bar{x},\bar{x}')=V(\vert \bar{x} - \bar{x}' \vert)=\frac{4\pi \hbar^2 a_s}{m} \delta(\vert \bar{x}-\bar{x}' \vert). 
\label{interac}
\end{equation}

Here and henceforth we identify the transition amplitude elements of equation (\ref{interac}) as $V_{\vec{i}\vec{j}\vec{ k}\vec{l}}$: 

\begin{equation}
\langle \vec{i}\vec{ j} \vert V \vert \vec{k}\vec{l} \rangle= g\int d^3 \bar{x} {\text {  }}  \varphi_{\vec{i}} (\bar{x})  \varphi_{\vec{j}} (\bar{x})  \varphi_{\vec{k}} (\bar{x})  \varphi_{\vec{l}} (\bar{x})=gV_{\vec{i} \vec{j}\vec{ k}\vec{l}},
\label{tae}
\end{equation}
where $g=4\pi \hbar^2 a_s/m $.

Our starting point is the derivation, in the weak coupling limit and the Markov approximation, of an irreversible master equation for the one body reduced density matrix operator\cite{Romero-Rochin,Brenig,Benedetto}. This is the quantum Boltzmann equation for the operator $\Gamma (\vec{i},\vec{j},t)$ that we define below. 

As it is well known, the essence of a Boltzmann equation is to consider that collisions among the particles are responsible for driving the system towards thermal equilibrium. Once in equilibrium every thermodynamic property depends on single particle properties only. Having in mind this idea, we write first the expectation value of any $N$-body single particle operator $O_1^N$ in the occupation number representation:

\begin{equation}
\langle O_1^N (t)\rangle = \text{Tr}  \text{  } O_1^N \rho (t) = \sum_{\{n_{\vec{m}} \}} \langle \{ n_{\vec{m}} \} \vert \sum_{\vec{i} ,\vec{j}}   a^{\dag}_{\vec{i}} a_{\vec{j}} \langle \vec{i} \vert O_1 \vert\vec{j} \rangle \rho (t) \vert \{ n_{\vec{m}} \} \rangle,
\end{equation} 
where $O_1$ is a one-body operator and $\rho(t)$ is the density matrix of the whole system. The occupation numbers $n_{\vec{m}}$ satisfy $\sum_{\vec{m}} n_{\vec{m}} =N$. From this expression we can identify the  one body reduced density matrix operator $\Gamma_{\vec{i}\vec{j}}(t)$:

\begin{equation}
\langle O_1^N (t) \rangle  = \sum_{\vec{i}\vec{j}} \langle \vec{i} \vert O_1 \vert \vec{j} \rangle \Gamma_{\vec{i}\vec{j}} (t), 
\end{equation} 
where 

\begin{equation}
\Gamma_{\vec{i}\vec{j}}(t)=\sum_{\{n_m \}} \langle \{ n_m \} \vert  a^{\dag}_{\vec{i}} a_{\vec{j}}\rho (t) \vert \{ n_m \} \rangle.
\label{gamma}
\end{equation}

We notice that by setting $\vec{i}=\vec{j}$ the one body reduced density matrix operator becomes the average occupation number in the state $\vec{i}$: 

\begin{equation}
\Gamma_{\vec{i}\vec{i}}(t)= \text{Tr} \text{  } a^\dag_{\vec{i}} a_{\vec{i}}  \rho(t) = \langle n_{\vec{i} }(t) \rangle.
\end{equation}
In order to derive the quantum kinetic master equation for $\Gamma_{\vec{i}\vec{j}}(t)$, and in particular the equation for the average occupation number $\langle n_{\vec{i}} (t) \rangle$, we start by considering the Schr\"odinger equation for the full density matrix of the system,

\begin{equation}
\frac{\partial \rho (t)}{\partial t} = -\frac{i}{\hbar} \left [ H , \rho (t) \right ].
\label{ecrho}
\end{equation}
After we substitute this equation into the resulting expression for the time derivative of equation (\ref{gamma}) we obtain:

\begin{equation}
\frac{\partial \Gamma_{\vec{i}\vec{j}} (t)}{\partial t} = \frac {i}{\hbar} \text{Tr } \{ \rho (t) \left[ H, \gamma_{\vec{i} \vec{j}} \right] \},
\label{qkme}
\end{equation} 
where we have defined $\gamma_{\vec{i} \vec{j}}= a^\dag_{\vec{i}} a_{\vec{j}}$. 

Then, we substitute $H= H_0 +V$ in equation (\ref{qkme}) and we obtain, 

\begin{equation}
\frac{\partial \Gamma_{\vec{i} \vec{j}}}{\partial t}= \frac{i}{\hbar} \text{Tr  } \{  \rho (t) \left [ H_0, \gamma_{\vec{i} \vec{j}} \right ] \} + \frac{i}{\hbar} \text{Tr  } \{  \rho (t) \left [ V, \gamma_{\vec{i} \vec{j}} \right ] \}.
\label{qkme2}
\end{equation}

To find an expression for the density matrix $\rho (t)$ we work in the Born approximation, where the perturbative character of the interaction term $V$ is considered. By keeping the first order of the interaction potential for $\rho(t)$ we get:

\begin{equation}
\rho(t)= \rho_G (t) -\frac{i}{\hbar} \int_0^t d\tau \left[ e^{-\frac{i}{\hbar} H_0\tau} V e^{\frac{i}{\hbar} H_0\tau} , \rho_G (t) \right], 
\end{equation}
with $\rho_G (t) = e^{-\frac{i}{\hbar} H_0 t } \rho(0) e^{\frac{i}{\hbar} H_0 t } $.

Now, we insert this last expression for $\rho(t)$ into the quantum kinetic equation (\ref{qkme2}) and we obtain:

\begin{eqnarray}
\frac{\partial \Gamma_{\vec{i} \vec{j}} (t)}{\partial t}&=& \frac{i}{\hbar} \text{Tr  } \rho_G (t) \left[ H_0, \gamma_{\vec{i} \vec{j}} \right]+ \frac{i}{\hbar} \text{Tr  } \rho_G (t) \left[ V, \gamma_{\vec{i} \vec{j}} \right]
\nonumber\\
&+&\left(\frac{i}{\hbar}\right)^2 \int_0^t d\tau \text{Tr  } \{  \rho_G (t) \left [e^{-\frac{i}{\hbar}H_0\tau} Ve^{\frac{i}{\hbar}H_0\tau}  ,\left[H_0,\gamma_{\vec{i} \vec{j}} \right] \right]\}
 \\&+& \left(\frac{i}{\hbar}\right)^2 \int_0^t d\tau \text{Tr  } \{  \rho_G (t) \left[e^{-\frac{i}{\hbar}H_0\tau} Ve^{\frac{i}{\hbar}H_0\tau}  ,\left[V,\gamma_{\vec{i} \vec{j}} \right] \right]\} .\nonumber
\end{eqnarray}

In order to have the final form of the quantum kinetic master equation, we should take into account the Markov approximation which consists in neglecting memory effects in the interacting kernel. Such an approximation is directly related to the time scales involved in the collision integrals. In other words, we must compare the time scale during the relaxation process against the correlation time arising from the duration of a single collision. Such comparison leads us to analyze the collision integrals on the right-hand side of the last equation. The elastic collision rate in $^{87}$Rb is approximately ten per second \cite{Cornell,Holland}. Let us denote by $\tau_0$ such an elastic collision rate. Since the correlation time involved in a collision is much shorter than the time scale on which the system relaxes to equilibrium, we can extent the upper limit of the collision integrals. That is, if $t \gg \tau_0$,

\begin{equation}
\int_0^t d\tau \rightarrow \int_0^\infty d\tau.
\end{equation}

Thus, the kinetic equation takes the form:
\begin{eqnarray}\label{eccin-t}
\frac{\partial \Gamma_{\vec{i} \vec{j}} (t)}{\partial t}&=& \frac{i}{\hbar} \text{Tr  } \rho_G (t) \left[ H_0, \gamma_{\vec{i} \vec{j}} \right]+ \frac{i}{\hbar} \text{Tr  } \rho_G (t) \left[ V, \gamma_{\vec{i} \vec{j}} \right]
\nonumber\\
&+&\left(\frac{i}{\hbar}\right)^2 \int_0^{\infty} d\tau \text{Tr  } \{  \rho_G (t) \left [e^{-\frac{i}{\hbar}H_0\tau} Ve^{\frac{i}{\hbar}H_0\tau}  ,\left[H_0,\gamma_{\vec{i} \vec{j}} \right] \right]\}\nonumber\\
 &+& \left(\frac{i}{\hbar}\right)^2 \int_0^{\infty} d\tau \text{Tr  } \{  \rho_G (t) \left[e^{-\frac{i}{\hbar}H_0\tau} Ve^{\frac{i}{\hbar}H_0\tau}  ,\left[V,\gamma_{\vec{i} \vec{j}} \right] \right]\}.
\end{eqnarray}
This is the irreversible master equation for the one body reduced density matrix operator $\Gamma_{\vec{i} \vec{j} }(t)$  from which collective phenomena that include the formation of vortex states and
collective excitations can be treated\cite{Williams,Abo-Shaeer,Madison,Edwards,Jin,Drummond}. Here we restrict our attention to the evolution in time of the average occupation number $n_{\vec{i}}$, that is, we consider $\vec{i} = \vec{j}$ in equation (\ref{eccin-t})

\begin{eqnarray}\label{maineq}
\frac{\partial \langle n_{\vec{i}}(t) \rangle}{ \partial t}&=&\frac{i}{\hbar} \text {Tr  } \rho_G(t) \left[ V,\gamma_{\vec{i} \vec{i}} \right]
\nonumber\\
&+&\left(\frac{i}{\hbar}\right)^2 \int_0^{\infty} d\tau \text{Tr  } \{  \rho_G (t)\nonumber\\
& \times& \left[e^{-\frac{i}{\hbar}H_0\tau} Ve^{\frac{i}{\hbar}H_0\tau}  ,\left[V,\gamma_{\vec{i} \vec{i}} \right] \right]\}.
\end{eqnarray}

To evaluate the trace on the right-hand side of equation (\ref{maineq}) we use Wick's theorem to separate each collision term into products of one body reduced density matrix operators. Then, after we substitute both, the ideal and interaction contributions given in (\ref{ideal}) and (\ref{interac}) respectively, and consider the summation over repeated indices we find:

\begin{eqnarray} \label{eccin}
\frac{\partial  n_{\vec{\eta}}(\tau)}{ \partial \tau}&=&-\sum_{\vec{i} \vec{k}\vec{ l}} \{ V_{\vec{i} \vec{\eta}\vec{ k} \vec{l}}^2 \text{ } \delta_{ \epsilon_{\vec{\eta}} +\epsilon_{\vec{i}} ,\epsilon_{\vec{k}} +\epsilon_{\vec{l}} }\nonumber\\
&\times & \left[  n_{\vec{l}} n_{\vec{k}} \left(1+ n_{\vec{\eta}} + n_{\vec{i}} \right)-  n_{\vec{i}} n_{\vec{\eta}} \left(1+ n_{\vec{k}} + n_{\vec{l}} \right) \right] \nonumber\\
&+& 2 \text{ }V_{\vec{i} \vec{ \eta} \vec{l} \vec{i}}  V_{\vec{k} \vec{l} \vec{k} \vec{ \eta}}  \text{ }\delta_{ \epsilon_{\vec{\eta}} -\epsilon_{\vec{l}} } n_{\vec{i}} n_{\vec{k}} \left( n_{\vec{l}} - n_{\vec{\eta}} \right) \},
\end{eqnarray}
where  $\tau=(8 a^2 m\omega^2/\hbar)t$ is a dimensionless time. 
The terms in the sum on the right-hand side express the energy conservation through the Kronecker delta-functions. Since we are dealing with a closed system the total energy $E$ and the total number of particles $N$ must be conserved.  By a simple substitution one can show that the stationary solution of equation (\ref{eccin}) for both, the ground state, and the excited states, is the Bose-Einstein distribution

\begin{equation}
n_{\vec{\eta}}=\frac{1}{\exp\left[\left(\epsilon_{\vec{\eta}}-\mu\right)/kT\right]-1},
\label{BED}
\end{equation}
where the temperature $\beta=1/kT$ and the chemical potential $\mu$ can be obtained by imposing  the  associated normalization conditions on the total energy $E=\sum_{\vec{\eta}}  \epsilon_{\vec{\eta}} n_{\vec{\eta}}$, and on the total number of particles $N=\sum_{\vec{\eta}} n_{\vec{\eta}}$.  However, if we consider $\mu =0$ in the Bose-Einstein distribution (\ref{BED}) and substitute it in on the right-hand side of equation (\ref{eccin}), we find that the left-hand side is zero if and only if $T=0$. In other words, for finite temperatures $T$, the ideal distribution for the excited states
\begin{equation}
n_{\vec{\eta}}^{\mu=0}=\frac{1}{\exp\left[\epsilon_{\vec{\eta}}/kT\right]-1},
\label{bose-ex1}
\end{equation}
 is not the stationary solution. This result is inconsistent with the fact that a non-uniform ideal Bose gases can exhibit the Bose-Einstein transition in two and three dimensions. Therefore, equation (\ref{eccin}) can only be correct for $T>T_c$ where $\mu<0$. Here, we should emphasize that BEC transition occurs in an ideal gas at at $\mu=0$ since the ground state energy $\epsilon_0=\frac{3}{2}\hbar\omega$ has been neglected.

From a theoretical point of view, the occurrence of the transition to Bose-Einstein condensation in an ideal gas depends on both, dimensionality and the confining potential where the atoms move \cite{Bagnato}. The combined aspects of dimensionality and isotropy, mathematically expressed in the density of states, determine if BEC can occur in an ideal gas \cite{Bagnato}. A 2D ideal uniform Bose gas does not exhibit BEC, by contrast to the 2D harmonically trapped one. Regarding the 3D case, both the homogeneous and the harmonically trapped gas can have the transition. However, on the experimental side where interactions participate, the ultracold atomic gases require to be confined in an external potential. 

Since in the present work we are dealing with a harmonic potential, the ideal distribution (\ref{bose-ex1}) should be the stationary solution for the excited states below $T_c$ in 2D and 3D cases, that is, when chemical potential $\mu$ becomes zero. As mentioned above, however, this does not occur. At equilibrium, the ideal ditribution $n_{\vec{\eta}}$ given by (\ref{bose-ex1}) does not satisfy Eq. (\ref{eccin}) for particles in the excited states unless $T=0$. This result suggests that equation (\ref{eccin}) is unable to describe the dynamics for temperatures below the critical temperature $T_c$. Therefore, taking into account the underlying physics, an alternative kinetic equation for $\mu =0$ and $T < T_c$ must be derived to properly describe the time evolution and equilibration of the gas.

\subsection{Kinetic theory for temperatures below $T_c$}
\label{sectionB}

The most important property of a dilute Bose gas that has experienced the transition to the condensate state is that as a result of having chemical potential equal to zero, the ground state becomes macroscopically populated. As a consequence of this fact the intensive thermodynamic variables, that adjust to the extensive ones, are exclusively determined by the particles occupying the excited states only. Perhaps a most relevant effect when a Bose gas has chemical potential equal to zero, is that the number of particles $N$ is no longer a thermodynamic variable. Although in the whole system the number of particles $N$ remains constant, the number of particles in the ground and excited states separately are not conserved since there is no energy cost in adding particles from the excited to the ground state and viceversa since $\epsilon_{i}=0$. Therefore, if one looks at the isolated fractions of particles in the ground and excited states, a continuous process of creation and annihilation of particles (having $\epsilon_{i}=0$) takes place in each one of them.

The physical process exposed above has not been considered previously, neither in the treatments involving kinetic descriptions, nor in the present one depicted in section \ref{sectionA}.  That is, for temperatures below $T_c$ the non conservation of particles in the separated fractions of ground and excited states has not been considered properly in the derivation of the kinetic equations describing the relaxation to the equilibrium state. 

The binary collisions are the key element where the non conservation of particles can be taken into account. Since such collisions are the elementary mechanism to reach thermal equilibrium and they involve particles in the ground and excited states, we must have special care in each collision event. Up to now, each pair of particles coming from either the ground state, or the excited states, transform after the collision into a pair of particles that again can occupy the ground state or the excited states, in such a way that there is no distinction between the fractions of particles in the ground and excited states respectively. However, if we consider the fact stated in the above paragraph regarding the non conservation of particles in the separate ground and excited states, and that the occupation number in the ground state becomes macroscopic $N_0 \sim N$, we can proceed in a different way treating separately the ground state and the excited states. In particular, we can explicitly assume that below the critical temperature, the fraction of the particles in the excited states participating in the collision process is not conserved. This procedure lead us to derive a new kinetic equation for the excited states. Then, by considering the fact that the {\it total} number of particles is conserved we can obtain the number of particles in the ground state by subtracting the particles in the excited states from the total number of particles $N$.

To model a process in which the number of particles in the excited states is not conserved, we first rewrite the interaction term given in equation (\ref{interac}) including separately the ground and excited states:

\begin{eqnarray}\label{full-int}
V&=& V_{{\vec 0}{\vec 0}{\vec 0}{\vec 0}}a_{\vec 0} ^\dag a_{\vec 0} ^\dag a_{\vec 0} a_{\vec 0} +
 \sum_{{\vec j} {\vec i} {\vec k} {\vec l} \neq 0}V_{{\vec j} {\vec i} {\vec k} {\vec l} } a_ {\vec j} ^\dag a_{\vec i} ^\dag a_{\vec k} a_{\vec l} \nonumber\\
&+&2\sum_{{\vec  j} {\vec k} {\vec l } \neq 0}V_{{\vec j} {\vec 0} {\vec k} {\vec l}}a_{\vec j} ^\dag a_{\vec 0} ^\dag a_{\vec k} a_{\vec l} +2 \sum_{{\vec i} {\vec j} {\vec k} \neq 0}V_{{\vec j} {\vec i} {\vec k} {\vec 0}}a_{\vec j} ^\dag a_{\vec i} ^\dag
a_{\vec k} a_{\vec 0} \nonumber\\
&+& \sum_{{\vec i} {\vec j}\neq 0}V_{{\vec j}{\vec i} {\vec 0} {\vec 0}}a_{\vec j} ^\dag a_{\vec i} ^\dag a_{\vec 0} a_{\vec 0} + \sum_{{\vec k}{\vec l}\neq 0} V_{{\vec 0} {\vec 0} {\vec k} {\vec l}} a_{\vec 0}^\dag a_{\vec 0}^\dag {\vec a_k} a_{\vec l} \nonumber\\
&+&4 \sum_{{\vec j} {\vec k} \neq 0}V_{{\vec j} {\vec 0} {\vec k} {\vec 0}}a_{\vec j} ^\dag a_{\vec 0}  ^\dag 
a_{\vec k} a_{\vec 0} +2 \sum_{{\vec j} \neq 0}V_{{\vec j} {\vec 0} {\vec 0}{\vec 0}}a_{\vec j} ^\dag a_{\vec 0} ^\dag a_{\vec k} a_{\vec 0}\nonumber\\
&+& 2\sum_{{\vec k} \neq 0}V_{{\vec 0} {\vec 0}{\vec k} {\vec 0}}a_{\vec 0} ^\dag a_{\vec 0} ^\dag a_{\vec k} a_{\vec 0},
\end{eqnarray}
where we have made explicit the fact that the ground state is not included in the sums. To take into account that the dynamics in the excited states (when $\mu=0$) is dictated with no constrains under conservation of the number of particles in such states, we propose to replace the interaction term given in equation (\ref{full-int}) by

\begin{eqnarray}\label{int-ex}
V= V_{{\vec 0}{\vec 0}{\vec 0}{\vec 0}} N_0^2 +\sum_{{\vec j}{\vec i}{\vec k}{\vec l} \neq 0} V_{{\vec j}{\vec i}{\vec k}{\vec l}}{\text {  }}a_{j} ^\dag a_{\vec i} ^\dag
a_{\vec k} a_{\vec l} + 2 \sqrt{N_0} \sum_{{\vec j} {\vec k} {\vec l} \neq 0} V_{{\vec j} {\vec k} {\vec l}{\vec 0}} \text {  }a_{\vec j} ^\dag
a_{\vec k} a_{\vec l}\nonumber\\
+ 2 \sqrt{N_0} \sum_{{\vec i}{\vec j}{\vec k} \neq 0} V_{{\vec j}{\vec i}{\vec k}{\vec 0}} \text{ } a_{\vec j} ^\dag a_{\vec i} ^\dag a_{\vec k} +
N_0\sum_{{\vec i}{\vec j} \neq 0} V_{{\vec j} {\vec i}{\vec 0}{\vec 0}} \text {  }a_{\vec j} ^\dag a_{\vec i} ^\dag +
N_0\sum_{{\vec k}{\vec l} \neq 0}V_{{\vec k} {\vec l}{\vec 0}{\vec 0}} \text {  }a_{\vec k}
a_{\vec l} \nonumber\\
+4 N_0  \sum_{{\vec j}{\vec k} \neq 0}V_{{\vec j}{\vec k}{\vec 0}{\vec 0}}\text {  }a_{\vec j} ^\dag  a_{\vec k} +2N_0^{3/2}
\sum_{{\vec j} \neq 0}V_{{\vec j}{\vec 0}{\vec 0}{\vec 0}}\text {  }a_{\vec j} ^\dag  +2 N_0^{3/2} \sum_{{\vec k} \neq 0}V_{{\vec k}{\vec 0}{\vec 0}{\vec 0}}\text {  } 
a_{\vec k}.
\end{eqnarray}
In this expression we have replaced the creation and annihilation operators $a_0^\dagger$ and $a_0$ by $\sqrt{N_0}$. This is the Bogoliubov approximation \cite{Bogolubov}. The aim behind such an assumption is to consider the macroscopic character of the particle population in the ground state $N_0$ for temperatures ranging below the critical temperature $T_c$. Taking the expectation value of any single-particle operator lead us obtain a quantity proportional to $N_0$.  

To justify our suggestion let us consider one of terms of the interaction potential (\ref{full-int}),

\begin{equation}
V=2\sum_{{\vec j}{\vec k}{\vec l}\neq 0}V_{{\vec j} {\vec 0}{\vec k}{\vec l}}a_{\vec j} ^\dag a_{\vec 0} ^\dag
a_{\vec k} a_{\vec l},
\end{equation}
This term represents the anhilation of particles in the states $\vec{k}$ and $\vec{l}$, and creation of particles in the states $\vec{j}$ and $\vec{0}$. If such a term is substituted by
\begin{equation}
V=2 \sqrt{N_0}\sum_{{\vec j}{\vec k}{\vec l}\neq 0}V_{{\vec j}{\vec k}{\vec l}{\vec 0}}a_{\vec j} ^\dag a_{\vec k} a_{\vec l},
\end{equation}
the number of particles in the excited states is explicitly not conserved, although of course, the missing particle belongs to the ground state. In other words, every interaction term in which the ground state participates is now omitted and the total number of particles in the system is conserved. After the expression (\ref{int-ex}) for the interaction potential is substituted in the master kinetic equation (\ref{maineq}) we obtain:

\begin{eqnarray} \label{eccin2}
&&\frac{\partial  n_{\vec{\eta}}(\tau)}{ \partial t}= -\sum_{{\vec i} {\vec k} {\vec l} \neq 0} \{ V_{{\vec i} {\vec \eta} {\vec k} {\vec l}}^2 \text{ } \delta_{ \epsilon_{\vec \eta} +\epsilon_{\vec i} ,\epsilon_{\vec k} +\epsilon_{\vec l} }\nonumber\\
&&\times  \left[  n_{\vec l} n_{\vec k} \left(1+ n_{\vec \eta} + n_{\vec i} \right)-  n_{\vec i} n_{\vec \eta} \left(1+ n_{\vec k} + n_{\vec l} \right) \right] \nonumber\\
&&+ 2 \text{ }V_{{\vec i} {\vec \eta} {\vec l} {\vec i}}  V_{{\vec k} {\vec l} {\vec k} {\vec \eta}}  \text{ }\delta_{ \epsilon_{\vec \eta} ,\epsilon_{\vec l} } n_{\vec i} n_{\vec k} \left( n_{\vec l} - n_{\vec \eta} \right) \} \nonumber\\
&&+n_{0}\sum_{{\vec k} {\vec i} \neq 0}V_{{\vec \eta}{\vec k}{\vec i} {\vec 0}} ^2\{ \text{  }
2\delta_{\epsilon_{\vec \eta}+\epsilon_{\vec i},\epsilon_{\vec k}}[n_{\vec k}(1+n_{\vec \eta}+n_{\vec i})-n_{{\vec i}}n_{\vec \eta}]\} \nonumber\\
 && +\delta_{\epsilon_{\vec \eta},\epsilon_{\vec k}
 +\epsilon_{\vec i}}[n_{\vec i} n_{\vec k}-n_{\vec \eta}(1+n_{\vec k}+n_{\vec i})]+ \nonumber \\
 &&+2n_{{\vec 0}}^2\sum_{\vec{k}}\delta_{\epsilon_{\vec \eta},\epsilon_{\vec k}
} V^2_{{\vec \eta}{\vec k}{\vec 0}{\vec 0}}(n_{\vec k}-n_{\vec \eta}) .
\end{eqnarray}
with
\begin{equation}
n_{0}(\tau)=N-\sum_{{\vec \eta}\neq 0}n_{{\vec \eta}}(\tau),
\end{equation}
being $n_0$ is the number of particles in the ground state. 

By substituting
\begin{equation}
n_{\vec{\eta}}=\frac{1}{\exp\left[\epsilon_{\vec{\eta}}/kT\right]-1},
\label{bose-ex}
\end{equation} 
one can show that this is the correct stationary solution of Eq. (\ref{eccin2}) for the excited states, that is, the Bose-Einstein distribution for zero chemical potential. Here the temperature $\beta=1/kT$  can be obtained by imposing  the  associated normalization condition on the total energy $E=\sum_{\eta \neq 0}  \epsilon_\eta n_\eta$. 

It is important to emphasize that Eq. (\ref{eccin2}) for $\mu = 0$ can be obtained in a direct way from Eq. (\ref{eccin}), for $ \mu < 0$, by separating the contributions of the ground and excited states, and considering that just those terms proportional to $n_0$ must be retained. That is, if in the separated sums of ground and excited states one neglects those terms in which the number of particles in the ground state does not appear, the derivation of the master equation is sensibly simplified.

\section{Simulation of the cooling process}
\label{section2}

In the preceding section we derived the kinetic equations that model a weakly interacting Bose gas confined in a harmonic potential under non-equilibrium conditions. These equations for the occupation numbers describe the thermalization process before  and after the occurrence of a BEC transition in the gas, Eqs. (\ref{eccin}) and (\ref{eccin2}) respectively. That is, for temperatures above the critical temperature $T_c$ where the chemical potential varies as temperature decreases, and for temperatures below $T_c$ where the chemical potential remains equal to zero.

In this section we present numerical results that model the cooling process as several cooling stages are performed, both for $T> T_c$ and $T< T_c$. While our goal is to show that the growth of a BEC can be described via successive thermalization processes, solving numerically equations for $\mu<0$ and $\mu=0$, (\ref{eccin}) and (\ref{eccin2}) respectively, we shall also point out along the way the difficulties posed by a finite-size calculation. From now on we use dimensionless units $\hbar=k_B=1$. The frequency $\omega$ will also be used as a dimensionless value, however, as it will be specified we shall compare different cases by keeping constant the so-called generalized density $N \omega^3$, as indicated by the thermodynamic limit in gases harmonically confined \cite{VRR-PRL05}.

\subsection{Numerical limitations}

The sums over the states on the right hand side in the kinetic equations for $\mu<0$ and $\mu=0$ (Eqs. (\ref{eccin}) and (\ref{eccin2})) extend to infinite, and therefore, an infinite number of coupled equations for the occupation numbers $n_{\vec \eta}$ must, in principle, be solved. In a real situation, however, prior to the first cooling stage, the gas has been precooled by some other methods and the atoms in the Bose gas do not occupy very high energy levels, namely,  the ratio of their occupation numbers compared to those in the lowest levels is negligible. A simple calculation shows that for $T \approx 10^{-6} K$ and a typical trap frequency $\omega \approx 2\pi (100)$ Hz, the maximum level occupied in the harmonic confinement corresponds to ${\mathcal N} \approx 100$ energy levels. In principle, considering such as the highest occupied level would be a good starting point for studying the dynamics of the condensation, and thus solving a finite number of equations. As we argue in the next paragraph, ${\mathcal N} = 100$ is still an unattainable situation. As we discuss below, we shall use ${\mathcal N} = 40$ as the maximum energy level and even this within the ``ergodic" approximation introduced in the following subsection.

There are two numerical tasks that become impracticable as the highest energetic level in the sums of kinetic equations grows. One is the number of coupled kinetic equations that have to be solved and the other is the determination of the collision integrals $V_{\vec{i}\vec{ j}\vec{ k}\vec{l}}$. Due to the degeneracy that grows with the energy level (see Eq. (\ref{degeneracy})), the number of coupled equations to be solved increases exponentially. For instance, for ${\mathcal N = 100}$ energy levels, there are $171,700$ states, while for ${\cal N} = 40$ one finds $11,480$ states, and thus, those are the number of non-linear coupled equations to solve in each case. On the other hand, since the right hand side of Eqs. (\ref{eccin}) and (\ref{eccin2}) involves an enormous number of sums, to be calculated at each time step, the calculation is even harder. For example, for ${\cal N} = 40$, the number of operations per iteration is of the order of $10^{15}$. We have estimated the time needed for this in a standard workstation, yielding of the order of $10^{7}$ seconds. This poses a very stringent restriction in the number of states (and thus, levels) that we can practically include. Therefore, we shall adopt the ``ergodic assumption" proposed in Ref. \cite{Holland} to deal with the dynamical description of the interacting Bose gas. 

The basic idea of the ergodic assumption is that the number of particles in a state $\vec m = (m_x,m_y,m_z)$ depends only on the energy of the state $\epsilon_{\vec m} = m\omega $, with $m = m_x +m_y+m_z$,  just as in thermal equilibrium. We then write $n_{\vec m} \equiv n_m$, namely asserting that states with the same energy have the same population. We recall that there are $g_m = (m+1)(m+2)/2$ states with the same energy. To implement the ergodic hypothesis, we can consider that the initial state is one of thermal equilibrium and then assume it remains ergodic throughout.

Using this hypothesis we  define an ergodic collision kernel \cite{Holland} as
\begin{equation}
I_{i k m l}=\delta_{\epsilon_i+\epsilon_k,\epsilon_m+\epsilon_k}\sum_{\vec{i},\vec{m},\vec{k},\vec{l}}V_{\vec{i}\vec{m}\vec{k}\vec{l}}^2\prod_{j}\delta_{\epsilon_{\vec{j}},\epsilon_{j}}
\label{kernel}1
\end{equation}  
with $j$ standing for all values $\{i, k,\eta, l  \}$. This collision kernel must be numerically obtained. Here we have calculated the kernel for $\epsilon_{j}\leq 40\omega$. The ergodic kinetic equation for $T>T_c$ is now,
\begin{eqnarray} 
\label{eccinergo}
g_{m}\frac{\partial  n_{{m}(\tau)}}{ \partial \tau}&=&-\sum_{i k l}  I_{i m k l}^2 \text{ } \delta_ { \epsilon_{m} +\epsilon_{i},\epsilon_{k} +\epsilon_{l} } \left[  n_{l} n_{k} \left(1+ n_{m} + n_{i} \right)-  n_{i} n_{m} \left(1+ n_{k} + n_{l} \right) \right] ,
\end{eqnarray}
while the kinetic equation for $T< T_c$ becomes,
\begin{eqnarray}
 \label{eccinergomu}
g_{m}\frac{\partial  n_{{m}(\tau)}}{ \partial \tau}&=&-\sum_{i k l}  I_{i m k l}^2 \text{ } \delta_ { \epsilon_{m} +\epsilon_{i},\epsilon_{k} +\epsilon_{l} } \left[  n_{l} n_{k} \left(1+ n_{m} + n_{i} \right)-  n_{i} n_{m} \left(1+ n_{k} + n_{l} \right) \right] +\nonumber\\
 &&+n_{0}\sum_{k i}I_{m k i 0}\{2\delta_{\epsilon_{m}+\epsilon_{i},\epsilon_{k}}[n_{k}(1+n_m+n_i)-n_i n_m]+\delta_{\epsilon_{m},\epsilon_{i}+\epsilon_{k}}[n_i n_k-n_\eta(1+n_k+n_i)]\}.\nonumber\\
&&n_{0}(\tau)=N-\sum_{m}n_m(\tau)
\end{eqnarray}
and we recall that the sums in the former case start with 0, while in the latter they begin with 1.

 Using the ergodic approximation lead us to reduce significantly the number of coupled equations. That is, for ${\cal N}$, one must solve only ${\cal N}$ non-linear coupled equations. Nevertheless, the ergodic collision kernel given by equation (\ref{kernel}) still remains as a numerical challenge since the number of terms in the sums grow exponentially. For ${\cal N} = 100$ levels the estimated time to calculate the ergodic kernel is approximately $\sim 9 \times 10^9$ seconds, while for ${\cal N} = 40$ it reduces to $\sim 2 \times 10^5$ seconds, a reasonable time.  To the best of our knowledge, $\mathcal N = 40$ levels for the calculation of the ergodic collision integrals $I_{i k \eta l}$ is the highest performed so far. We shall use this number of levels, within the ergodic hypothesis, to numerically simulate an energetic cooling protocol, as it is now described.

\subsection{Energetic cooling protocol}

\begin{figure}[htbp]
\begin{center}
\includegraphics[width=3.5in]{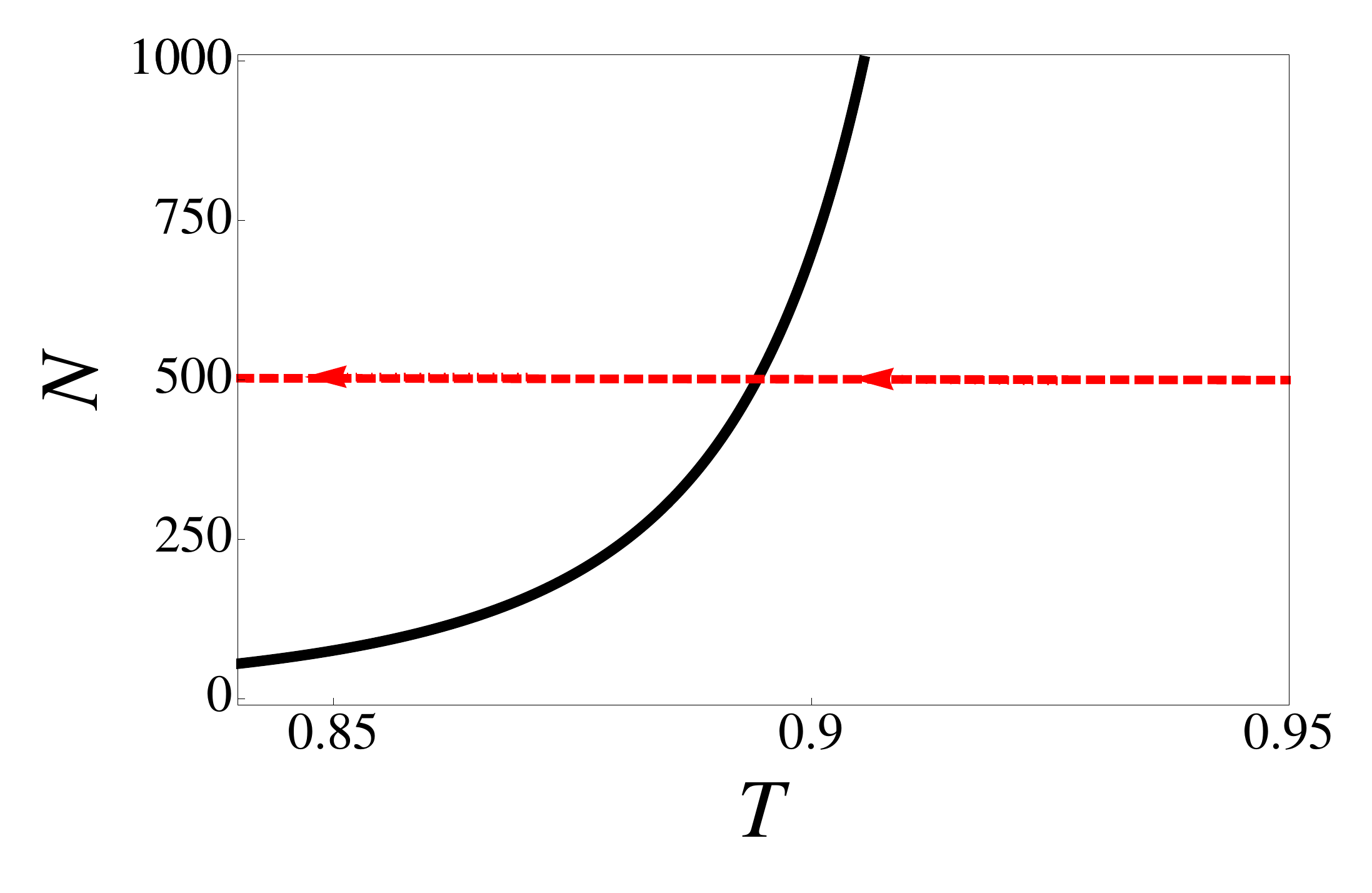} 
\end{center}
\caption{(Color online) Scheme of an energetic cooling protocol. The solid line indicates the dependence of the number of particles in the excited states a function of $T$ for $\mu=0$, for $N=500$, while the dashed horizontal line represents a cooling sequence at constant number of particles.} 
\label{Fig1}
\end{figure}

For our study we propose an energetic cooling protocol. This scheme consists in maintaining fixed the number of particles during the whole cooling process while several energetic cooling stages lead the system to decrease its temperature. The basic idea of cooling consists in transferring particles from higher energy levels to lower ones. 
The system, being then out of equilibrium, will evolve to the equilibrium state via the kinetic equations. Once in equilibrium with a lower temperature, the process of particle redistribution is performed several times to follow the evolution towards condensation. Fig.\ref{Fig1} illustrates, in a diagram $N$ vs $T$, a possible sequence of an energetic cooling process. The horizontal dashed (red) line shows a non-equilibrium process at constant number of particles that, starting from a temperature $T > T_c$, crosses the solid (black) line which represents the number of particles $N$ in the excited states as a function of the critical temperature $T_c$, namely, the onset of Bose-Einstein condensation.

An energetic cooling mechanism may be implemented via ``stimulated anti-Stokes Raman-like" processes, as described in Section I. The numerical simulation is initiated by distributing $N$ particles following the Bose-Einstein statistics in the first 40 levels. To perform the first energetic cooling step we remove the particles lying, say, in level $m$, and distribute them uniformly in the level $m-1$ for $m \le 40$. The system is allowed to evolve \cite{Euler} through equations (\ref{eccinergo}), for $\mu < 0$ which corresponds to $T < T_c$,  until it reaches the equilibrium state associated to this first cooling stage. This process is successively repeated to follow the transition towards the condensation transition. Once the transition is crossed, the dynamics should follow that of equations (\ref{eccinergomu}), corresponding to $\mu = 0$ and $T < T_c$. As we shall discuss in detail, due to the finite number of particles, there is a clear crossover between the solutions of the two ergodic sets of equations, rather than a sharp, discontinuous transition, as expected in the thermodynamic limit.

\begin{figure}[htbp]
\begin{center}
\includegraphics[width=3.5in]{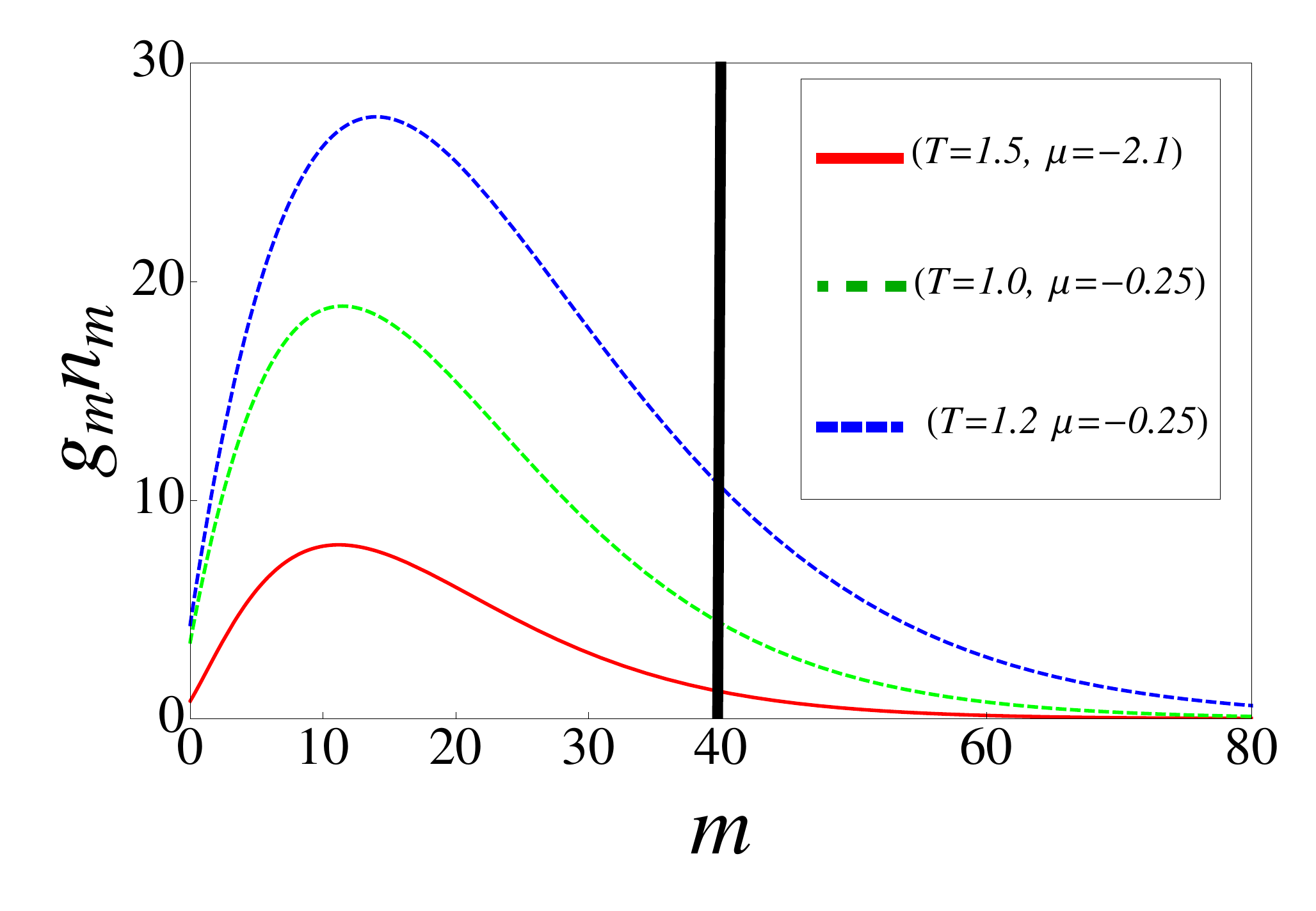} 
\end{center}
\caption{(Color online) Bose-Einstein distribution for different values of the chemical potential $\mu$ and temperature $T$. The vertical line indicates the maximum number of energy levels considered in our numerical simulations.} 
\label{Fig2}
\end{figure}

To make use of the energetic cooling protocol we prepare the initial state considering the maximum energy level allowed by our computational capabilities. To ensure that all the particles participate in the dynamics during each energetic cooling step, it is necessary to select initial appropriate values of $\mu$ and $T$. In Fig. \ref{Fig2} we plot the occupation number as a function of energy level in a harmonic oscillator for three pairs of values of $T$ and $\mu$, indicated in the figure. As it can be seen from this plot, the number of particles below the level ${\mathcal N}=40$ depends on the values of $T$ and $\mu$. Given a fixed values of $T$ and $\mu$, one can estimate the relative number of particles above ${\mathcal N}=40$  through

\begin{equation}
\Delta N= \frac{N_{\infty}(T, \mu)-N_{40}(T, \mu)}{N_{\infty}(T, \mu)}.
\end{equation}
with $N_{\cal N}(T, \mu)=\sum _{m =0}^{\cal N} g_m \frac{1}{e^{ \omega (m-\mu)/T}-1}$ and $N_{\infty}(T, \mu)$ the corresponding value of $N$ for an ``infinite" number of states ${\cal N} \to \infty$. 

For practical purposes we have selected two pairs of initial values of temperature $T$ and chemical potential $\mu$ to illustrate the predictions of the kinetic equations for $\mu < 0$ and $\mu = 0$ (equations (\ref{eccinergo}) and (\ref{eccinergomu})). These values are, ($T=1.5,\mu=-2.1$) and ($T=1.0,\mu=-0.25$). Such values of $T$ and $\mu$ ensure that less than 6 and 8 percent of particles respectively will remain above the ${\mathcal N}=40$ level. Thus, one can make certain that the majority of the particles will participate in the dynamics, at each cooling stage across the condensation transition. Up to ${\mathcal N}=40$ energy level the number of particles in each case is $N_{40}(1.5,-2.1)=100$ and $N_{40}(1.0,-0.25)=500$. The critical temperatures for $N = 100$ and $N=500$ are $T_c = 0.801$ and $T_c = 0.854$ respectively. 

First we proceed to analyze the predictions of equations (\ref{eccinergo}) for $N = 100$, namely for the initial state $T=1.5,\mu=-2.1$. In Fig. \ref{Fig3a} we plot the growth of the normalized number of particles in the ground state $n_0$ along the whole cooling protocol as a function of the dimensionless time $\tau$, with a time step of $5 \times 10^{-5}$ \cite{Euler,Numerics}. As it can be seen from figure \ref{Fig3a}, the growth of $n_0$ as a function of time is not a linear function. At earlier cooling stages there is an efficient cooling. Then, once the system enters into the condensation transition the growth rate of particles in the ground state decreases. This fact is a consequence of having most of the particles in the ground state and an almost empty population in the excited states. We notice that this is accompanied by the fact that in the latest stages the chemical potential is essentially zero; we further discuss this point below.

The cooling process modeled through the protocol allows us to follow the evolution of the temperature and chemical potential as consecutive energetic cooling stages are carried out. At each energetic cooling stage the gas thermalizes at well defined values of $\mu$ and $T$; these are obtained by fitting the Bose-Einstein distribution function to the corresponding stationary distributions $n_m$ at each stage, namely, the stationary values are fit with  
\begin{equation}
n_m= \frac{1}{e^{ \omega(m-\mu)/T} -1},
\label{BED}
\end{equation}
with $\mu < 0$ since we are following the dynamics across the transition with the kinetic equation (\ref{eccinergo}). In Fig. \ref{Fig3b} we plot the stationary distributions of the occupation numbers $n_m$. Different curves correspond to equilibrium values  $T$ and $\mu$, at  different cooling stages as indicated in the legend. The curves lying on the points correspond to the best Bose-Einstein distribution fit. 
\begin{figure}[h]
\subfigure[]{\includegraphics[width=0.4\textwidth]{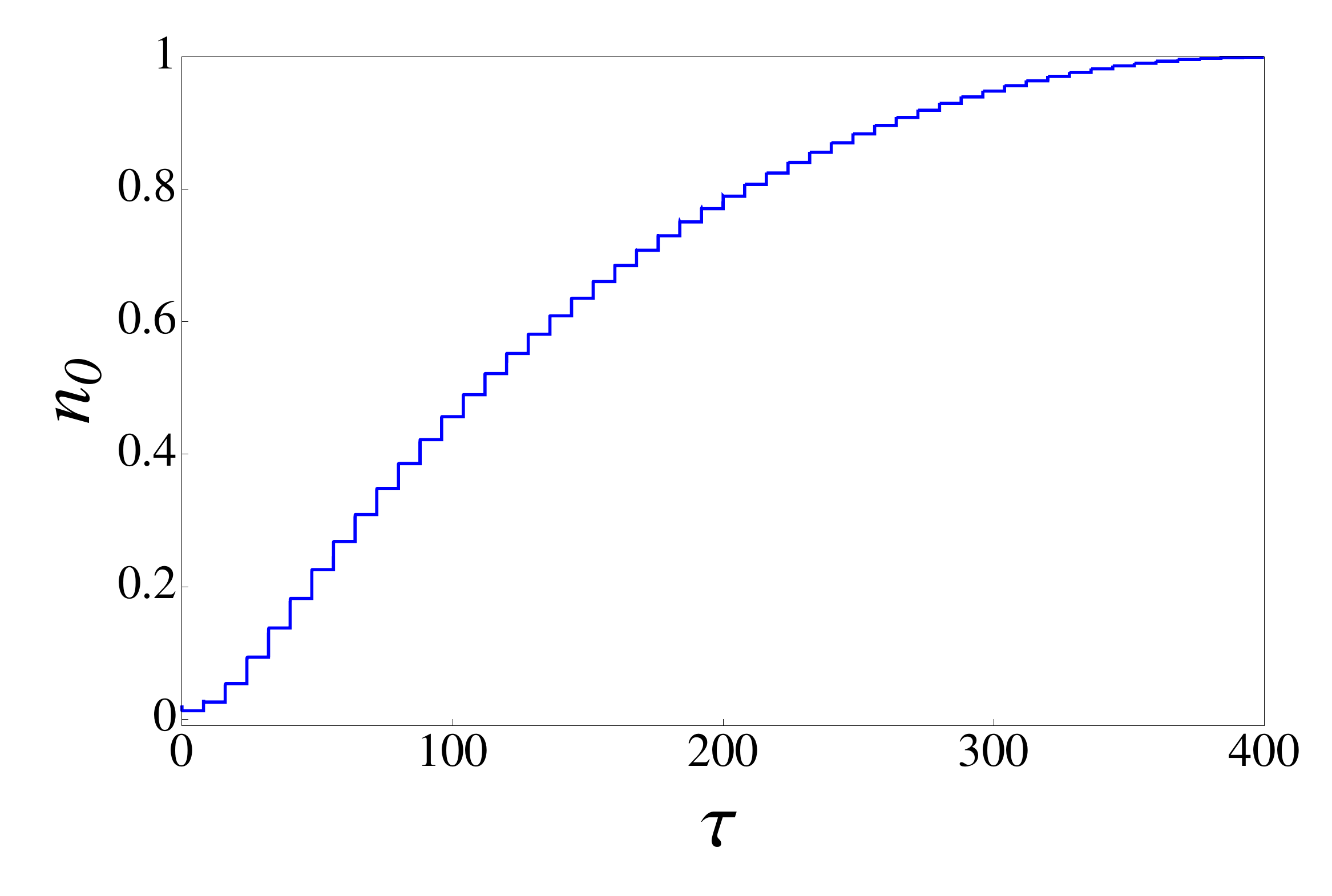}\label{Fig3a.pdf}}
\subfigure[]{\includegraphics[width=0.4\textwidth]{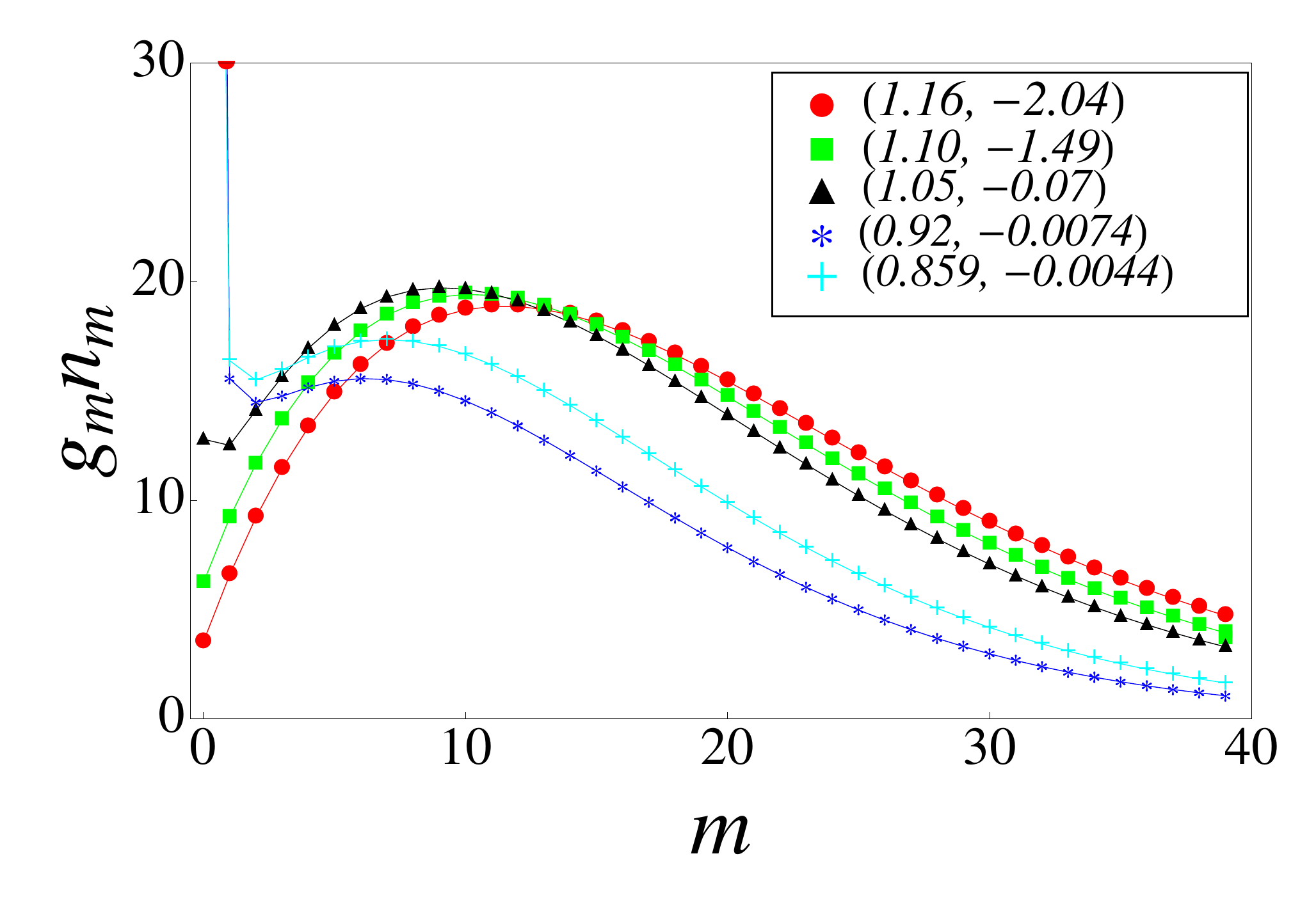}\label{Fig3b.pdf}}
\caption{(Color online) (a) Growth of the normalized number of particles in the ground state $n_0$ as a function of dimensionless time $\tau$ along the whole sequence of energetic cooling. (b) Stationary distributions of the ground and excited states for some energetic cooling steps. Different curves correspond to different values of $(T,\mu)$ as indicated in the labels. Numerical simulations in both figures were performed using equations (\ref{eccinergo}) with $N=100$. }
\end{figure}
As one can observe from this figure, as $T$ decreases and $\mu$ approaches zero from below, the number of particles in the condensate becomes the majority with respect to those associated to the excites states. As a matter of fact in figure \ref{Fig3b} we have omitted the stationary distributions of $n_0$ associated to $T=0.92$ and $\mu =-0.0074$ and $T= 0.859$ and $\mu= -0.0044$, since they would appear out of scale. Different curves in \ref{Fig3b} indicate the path towards the onset of degeneracy, that is, the behavior of the ground state occupation number as $\mu \rightarrow 0^-$. Two important facts we stress here, first, that around $T \approx 0.9$, BEC seems to develop; but second, due to finite, actually rather small, number of particles, the chemical potential does not become zero, although it does becomes very small. That is, it is not possible to observe a sharp transition. This can be further observed in Fig. \ref{Fig4} where we plot the chemical potential $\mu$ as a function of $T$, for $N=100$ and $N=500$ number of particles. From this figure one can observe how for $N=500$ the transition becomes more apparent and it should tend towards a true sharp transition in the thermodynamic limit $N \rightarrow \infty$ \cite{Yang}. This behavior may seem to indicate that, since the chemical potential does not become zero unless in the thermodynamic limit, the kinetic equations for $\mu < 0$ may suffice to describe the dynamics correctly, as done in Refs. \cite{Holland,Gardiner} for example. It is the purpose of this section to highlight the importance of the kinetic equations for $\mu =0$.

\begin{figure}[htbp]
\begin{center}
\includegraphics[width=3.5in]{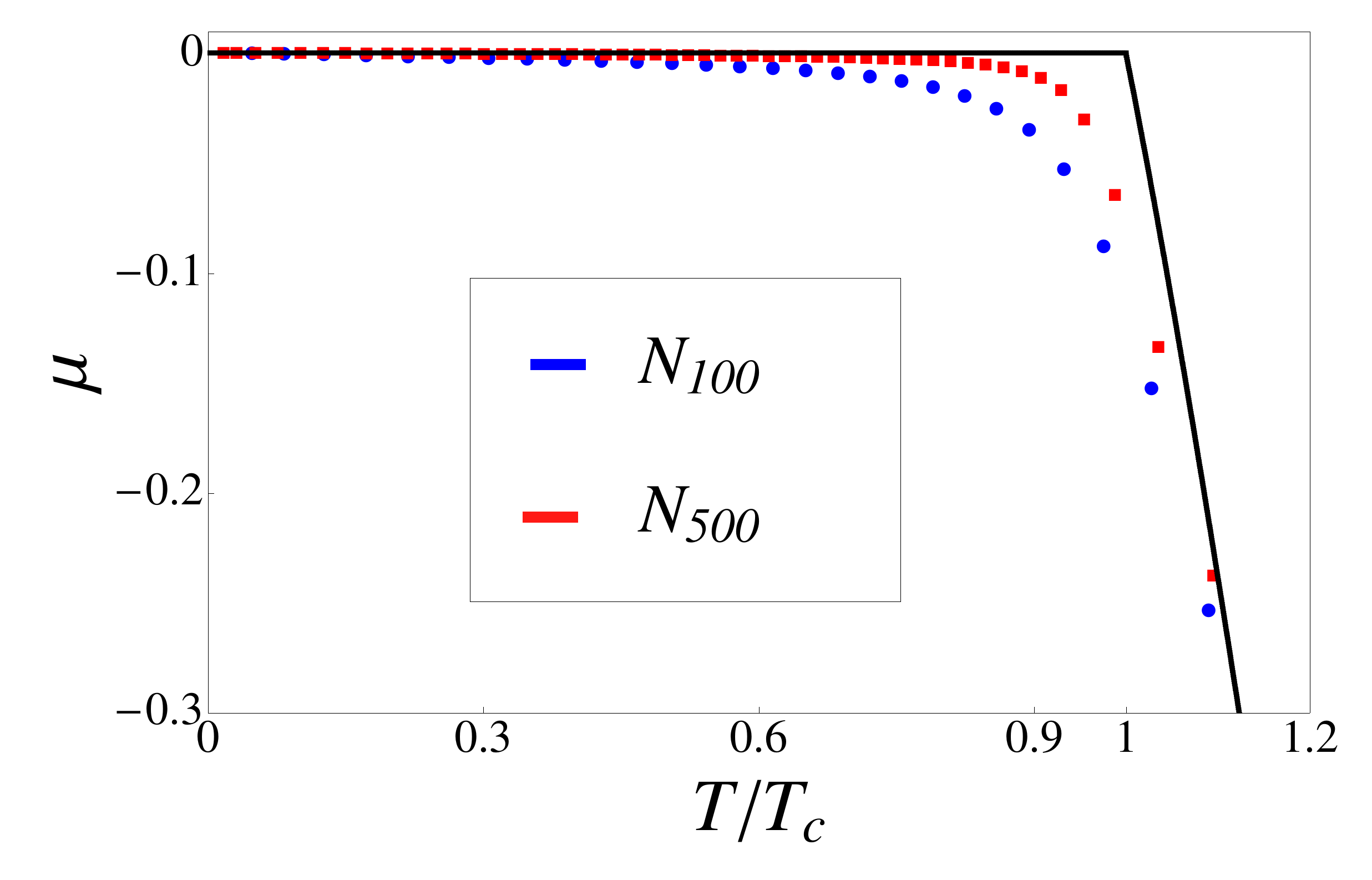} 
\end{center}
\caption{(Color online) Dimensionless chemical potential $\mu$ as a function of $T$ along the whole sequence of energetic cooling. Circles and squares are symbols that label $N=100$ and $N=500$ respectively for kinetic equation (\ref{eccinergo}) for $\mu<0$. Black solid line is associated to the thermodynamic limit.} 
\label{Fig4}
\end{figure}

To enquiry the predictions of kinetic equations for $\mu=0$ and then to contrast them with those associated to the equation for $\mu<0$ we proceed in an analogous way to that described above. That is, we simulate numerically the scheme of an energetic cooling protocol but now with equations (\ref{eccinergomu}) for $\mu=0$. We illustrate our numerical results for $N=500$ initiating at $T=1.0$ ($T_c = 0.854$). Since the chemical potential in this case is zero we obtain the value of $T$ by fitting to the stationary occupation numbers the ideal Bose-Einstein distribution for $\mu=0$,

\begin{equation}
n_m= \frac{1}{e^{\omega m/T} -1}.
\end{equation}
\begin{figure}[h]
\subfigure[]{\includegraphics[width=0.4\textwidth]{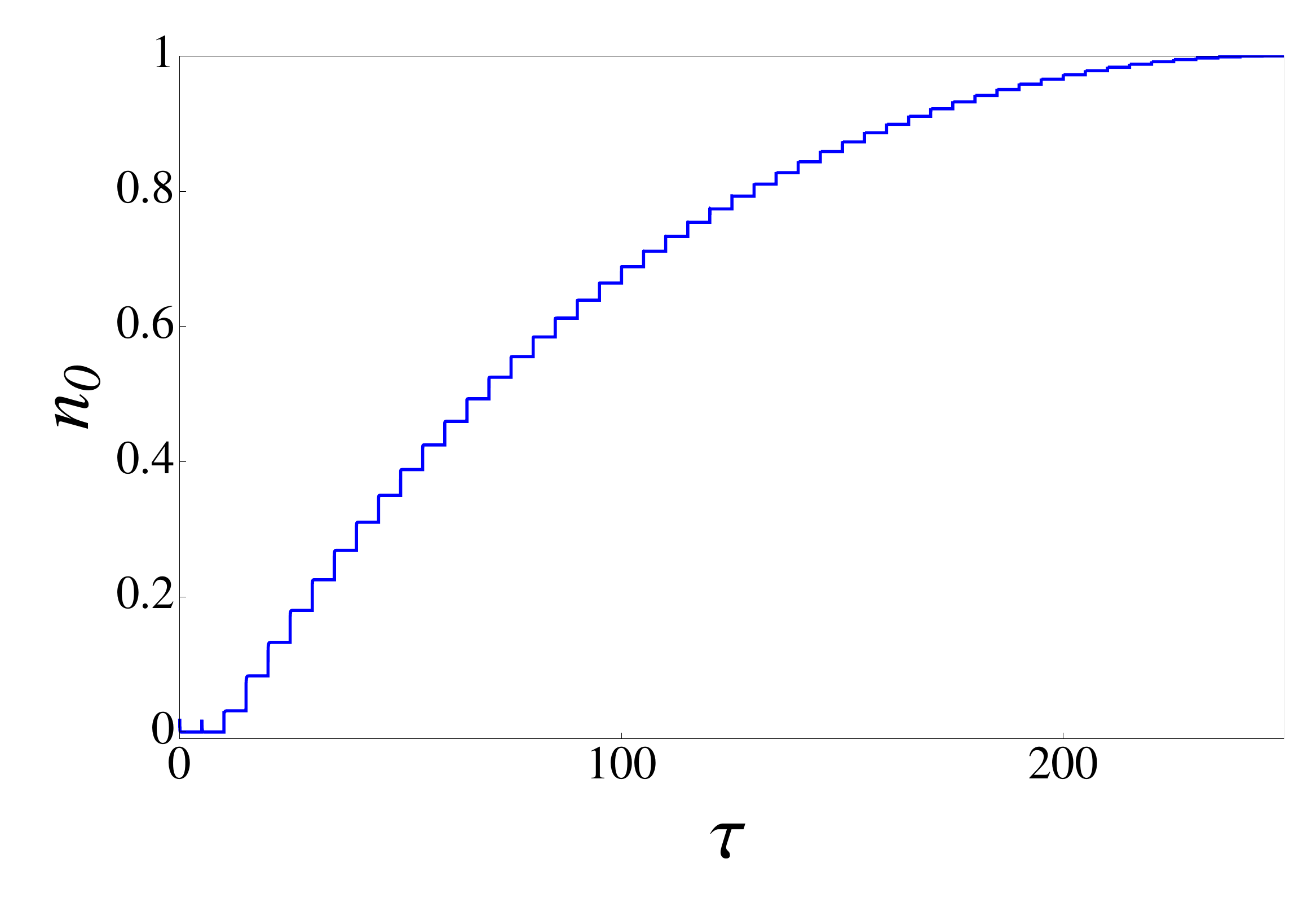}\label{Fig5a.pdf}}
\subfigure[]{\includegraphics[width=0.4\textwidth]{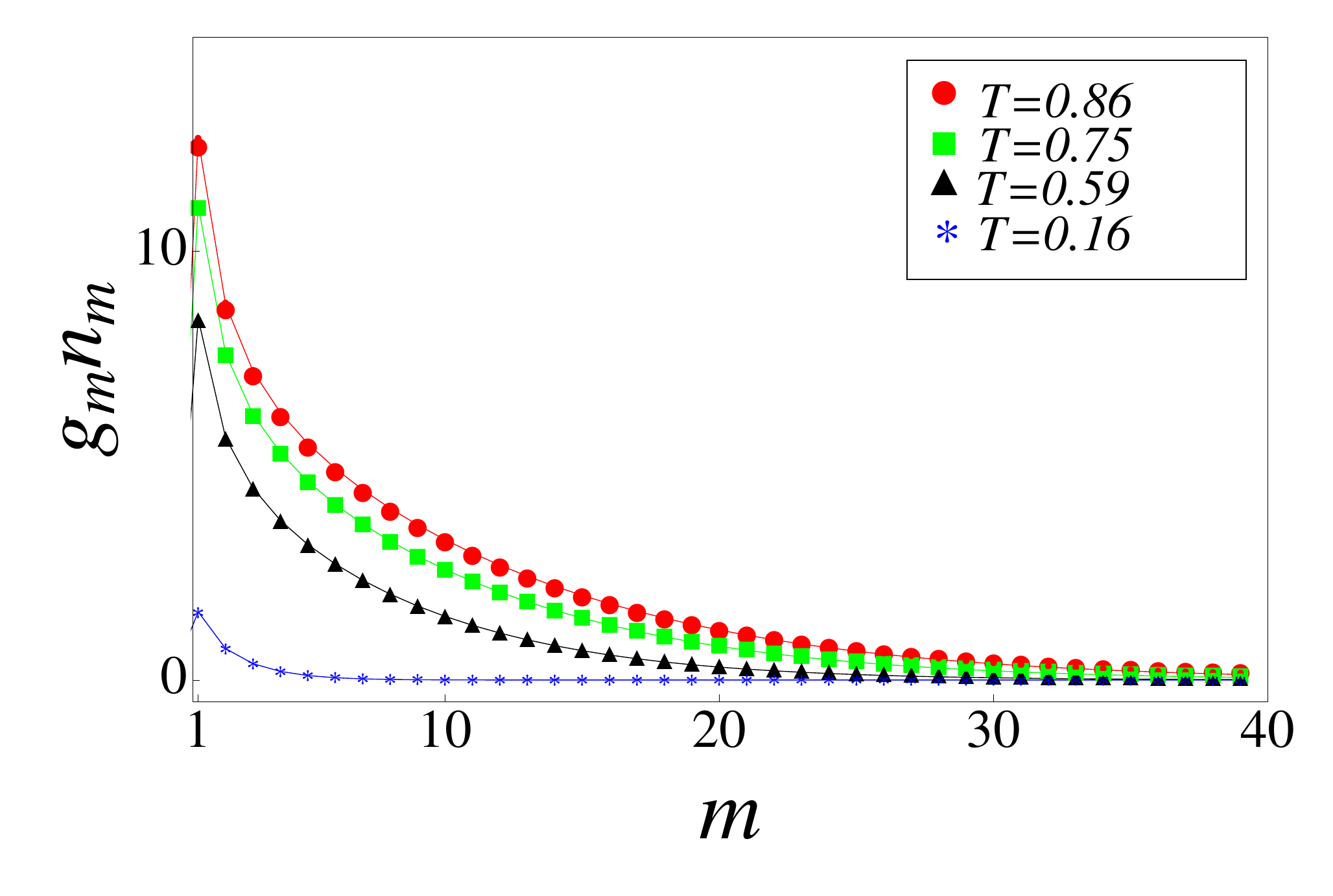}\label{Fig5b.pdf}}
\caption{(Color online) (a) Growth of the normalized number of particles in the ground state $n_0$ as a function of dimensionless time $\tau$ along the whole sequence of energetic cooling. (b) Stationary distributions of the excited states for some energetic cooling steps. Different curves correspond to different values of $(T,\mu)$ as indicated in the labels. Numerical simulations in both figures were performed using equations (\ref{eccinergomu}) with $N=100$. }
\end{figure}

In Fig. \ref{Fig5a} we plot the time evolution of the condensate $n_0$ during the whole energetic cooling process. Once more, at the start, the cooling is very efficient, but as $T \to 0$, it slows down since there are very few particles in the excited states. In Fig. \ref{Fig5b}, we plot the number distribution as a function of the energy level, for different temperatures. BEC is  clearly described for $T < T_c$ by using the corresponding equations (\ref{eccinergomu}) tailored for $\mu = 0$. This dynamics is very precise, less sensitive to finite size effects, and can be extended with no problem all the way to $T = 0$. We now compare the solutions both for $\mu <0$ with $\mu = 0$, in the light of finite size effects and the limited number of levels we can use.

In Figs. \ref{Fig6a} and \ref{Fig6b} we illustrate in a diagram $n$ vs  $T/T_c$ the stationary normalized number of particles in the ground and excited states for $N=500$ along the cooling process. Fig.  \ref{Fig6a} shows a cooling process starting at a temperature $T > T_c$ and going to a temperature as low as possible, using the kinetic equations for $\mu < 0$, Eq. (\ref{eccinergo}). Fig.  \ref{Fig6b} shows a cooling initiating at $T > T_c$ but using equations (\ref{eccinergomu}) valid for $\mu = 0$ only and, again, going towards $T = 0$. The curves in black are the results in the thermodynamic limit, showing the above mentioned sharp transition. Several comments are in order. First of all, the solution for $\mu < 0$, while the BEC transition is rather smooth, simply due to finite size effects, it appears that it can be extended to very low temperatures, near $T = 0$, where it is supposed no to be quite valid. This robustness is because the chemical potential can become very small, though not quite zero, see Fig. \ref{Fig4}. As we discuss below, the solution for $\mu < 0$ becomes numerically unstable as $T \to 0$. On the other hand, the kinetic equations for $\mu = 0$ become extremely good in the vicinity $T = 0$, notwithstanding the small number of particles. This is so, because ${\cal N} = 40$ energy levels suffices to describe the part of the gas in the excited states. As expected, this solution becomes less confident as $T \to T_c$ from below, and similarly to the equations for $\mu < 0$ near absolute zero, the numerical description for $\mu <0$ becomes unstable as BEC transition is approached. One expects that in the limit of very large number of particles, and large number of levels included in a numerical calculation, the limitations of both descriptions in the regions where are not supposed to be valid, should be more apparent. This is further discussed below.

Figs. \ref{Fig7a} and \ref{Fig7b} show details of calculations near $T = 0$ and $\mu < 0$, Eq. (\ref{eccinergo}), and for near $T_c$ and $\mu = 0$, Eq. (\ref{eccinergomu}), respectively. The purpose of these figures is to exemplify the numerical instabilities of the corresponding descriptions. Fig.  \ref{Fig7a} shows the number of particles in the condensate, with $\mu < 0$, with a very large number of particles, $N = 10^5$, which can be performed due to the very small value of the temperature. One sees, first, that the solution deviates from the that of the thermodynamic limit (solid line), then it becomes unstable. The simpler explanation is that, at these temperatures $\mu$ must be zero, but the fitted Bose-Einstein distribution given by Eq. (\ref{BED}) cannot take $\mu = 0$ for a divergence would occur. Although not shown here, the solution for $\mu = 0$ fits smoothly the solid curve all the way down to $T = 0$. On the other hand, Fig. \ref{Fig7b} shows the stationary total occupation number in the excited states for $\mu = 0$ and $T < T_c$. Similarly to the previous case, as $T_c$ is approached, a numerical difficulty arises in the form of a double valued solution. Again, this exhibits that this solution requires not only more particles but more equations for higher energy levels. It is satisfactory, nevertheless, to find that, apart from near the vicinity of $T_c$, the solution for $\mu = 0$, which explicitly takes the Bogoliubov approximation into account, fits extremely well the curves of the thermodynamic limit.

\begin{figure}[h]
\subfigure[]{\includegraphics[width=0.4\textwidth]{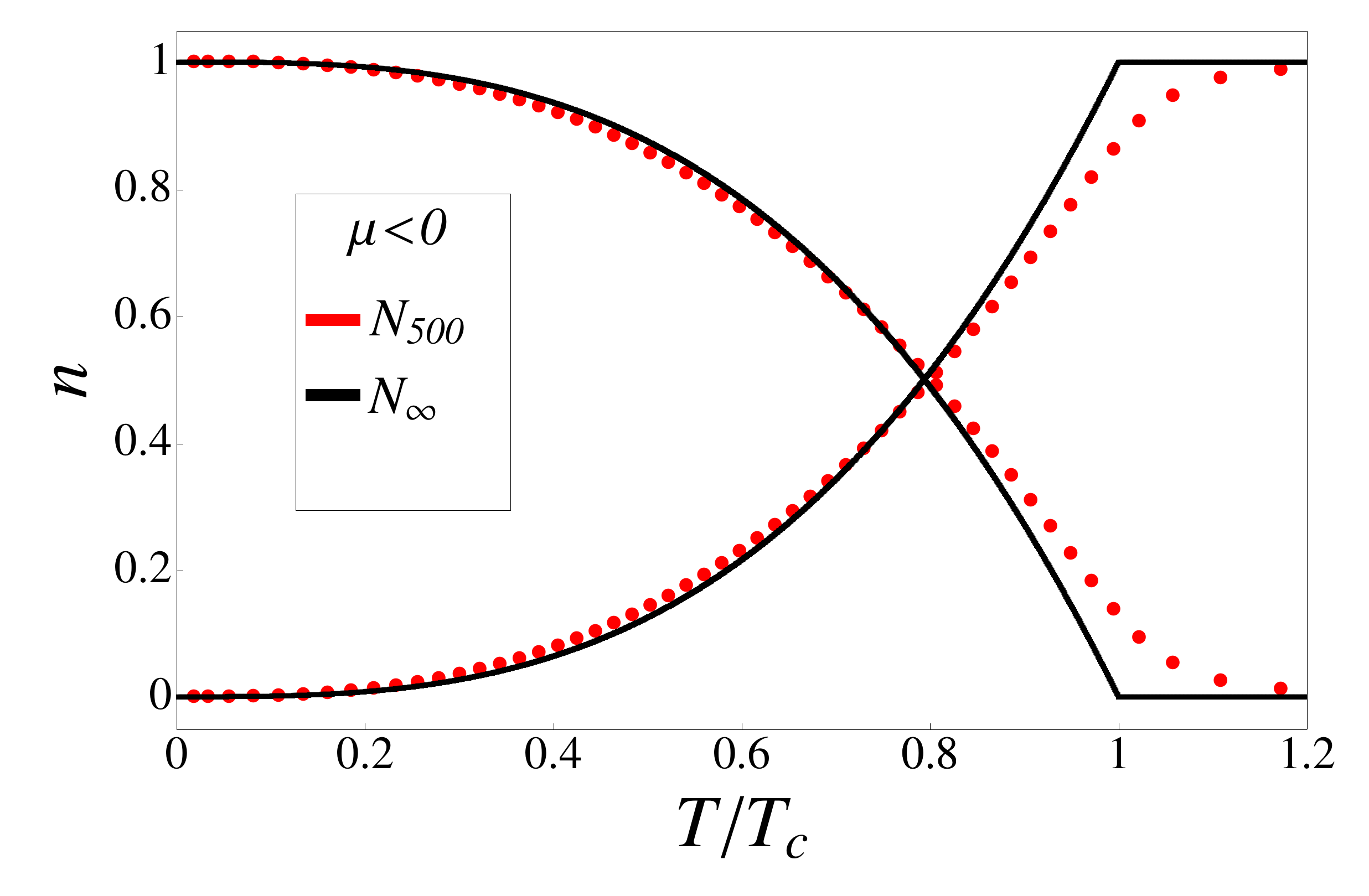}\label{Fig6a.pdf}}
\subfigure[]{\includegraphics[width=0.4\textwidth]{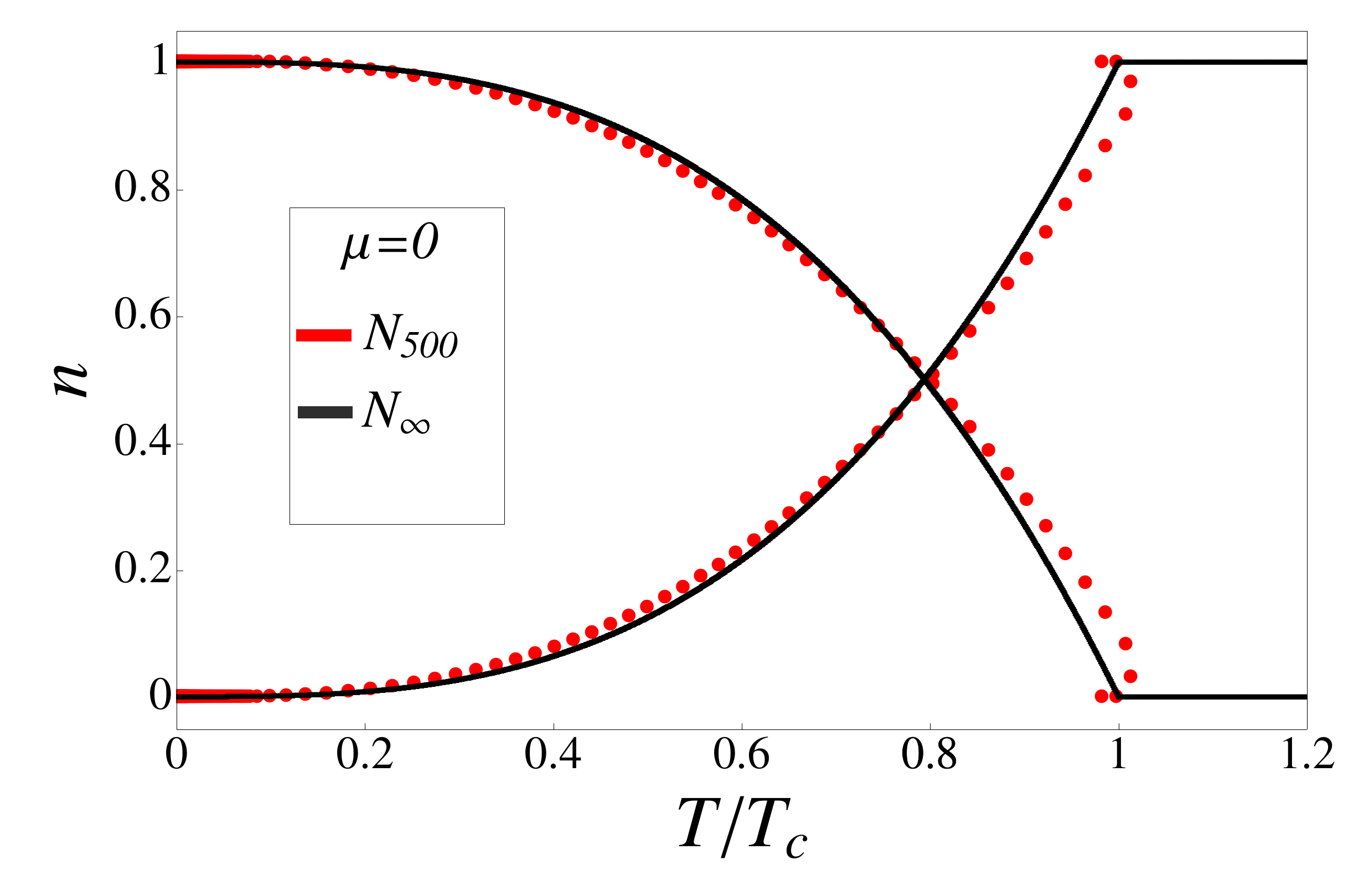}\label{Fig6b.pdf}}
\caption{(Color online) Diagrams of the normalized number of particles $n$ as a function of $T$ along the whole sequence of energetic cooling. Black solid line is associated to the thermodynamic limit. (a) and (b) correspond to equations (\ref{eccinergo}) and (\ref{eccinergomu}) respectively. Those simulations are associated to $N=500$.}
\end{figure}

\begin{figure}[h]
\subfigure[]{\includegraphics[width=0.4\textwidth]{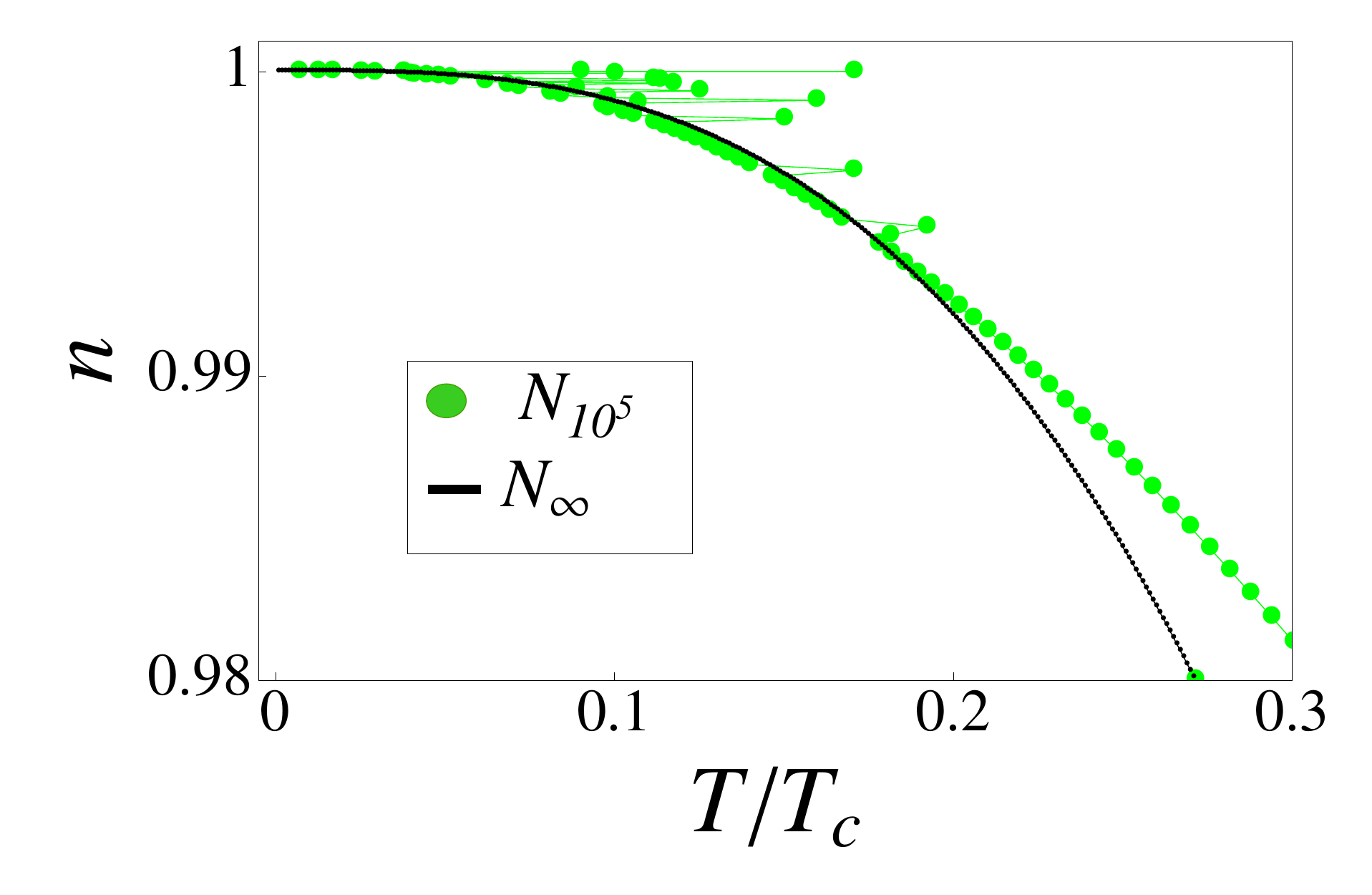}\label{Fig7a.pdf}}
\subfigure[]{\includegraphics[width=0.4\textwidth]{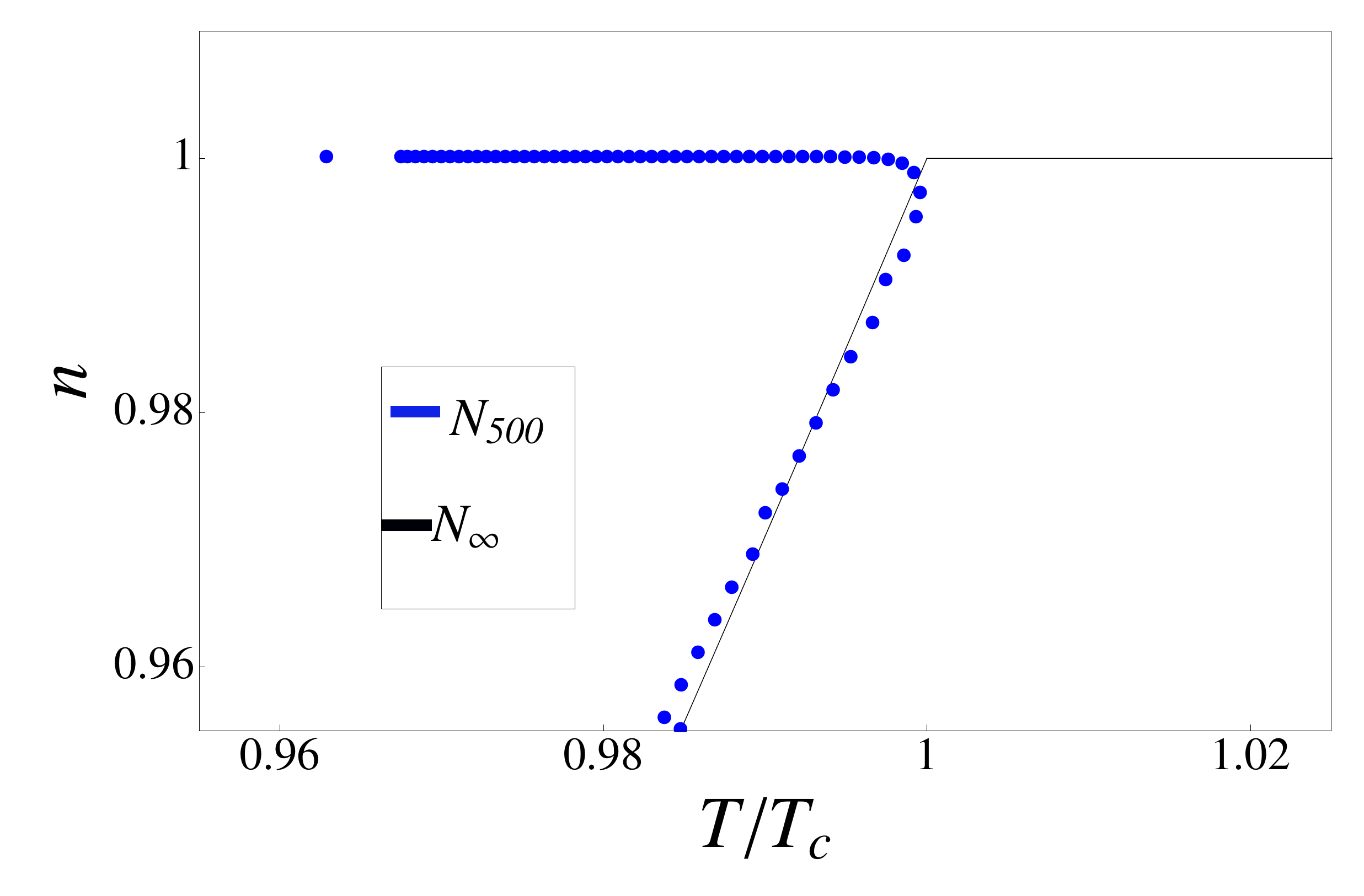}\label{Fig7b.pdf}}
\caption{(Color online) Diagrams of the normalized number of particles $n$ as a function of $T$ along a reduced stage of the energetic cooling. (a) corresponds to equations (\ref{eccinergo}) with $N=10^5$ and (b) corresponds to equations  (\ref{eccinergomu}) with $N=500$. The black solid  curve corresponds to the thermodynamic limit.} 
\label{Fig6}
\end{figure}

\section{Final remarks}
\label{section3}

We have studied the cooling process that leads a low-interacting Bose gas confined in a harmonic potential to reach the Bose-Einstein transition. For this purpose we modeled the reduction of temperature in the gas through an energetic cooling protocol. In an analogous way to the evaporative cooling mechanism, the fundamental principle behind such a process, is that the temperature in a gas can be reduced in several stages by removing selectively the most energetic particles, and then allowing the system to thermalize to an equilibrium state with a lower temperature. The essential mechanism that leads the low interacting gas to reach the equilibrium state are the inter particle collisions. Thus the path to equilibrium at each cooling step can be followed dynamically through a quantum kinetic Boltzmann description. We derived the equations that describe the relaxation to the equilibrium in the gas taking into account that the cooling process must be divided in two parts, one for temperatures above the critical temperature $T_c$  at which the gas is in the normal phase, and the other for temperatures below $T_c$ when the transition to Bose-Einstein condensation takes place.

To derive the irreversible master equations for the one body reduced density matrix operator in both cases ($T>T_c$ and $T<T_c$), we worked in the Born and Markov approximations that assume the perturbative character of the interaction term that represents the collisions among the particles, and that memory effects of the collision kernel are negligible, respectively. Among the dynamics of any one-body reduced density matrix operator $\Gamma_{i j}(t)$, we specifically followed the evolution in time of the occupation number operator $\langle n_{i}(t) \rangle=\Gamma_{i i}(t)$ to determine the particle population, in a single quantum state $i$, as a function of time during the thermalization process. 

First, based on the fact that Bose-Einstein condensation can occur in a harmonically trapped Bose gas in two and three dimensions, we derived for $T>T_c$ the kinetic equation that describes the relaxation to a stationary state of a collection of particles out of equilibrium, Eq. (\ref{eccin}). We demonstrated that such equation satisfies that the total energy and the total number of particles are conserved, and that the ideal Bose-Einstein distribution with $\mu \neq 0$ is the stationary solution of equation (\ref{eccin}) for both the ground state, and the excited states. However, since equation (\ref{eccin}) does not describe the dynamics for the ground and excited states when the chemical potential becomes zero or the temperature diminishes below $T_c$, we derived a new kinetic equation, taking into account the underlying physics using the Bogoliubov approximation \cite{Bogolubov}, to properly describe the evolution in time of a  low-interacting Bose gas for $T<T_c$.

The quantum kinetic equation for $T<T_c$ was derived considering the physical significance of the fact that the chemical potential remains unchanged at the occurrence of the Bose-Einstein transition. In other words, we took into account that the thermodynamic state of the system is solely determined by the particles occupying the excited states. And therefore, considering the fact that the number of particles $N$ is no longer a thermodynamic variable although the total number of particles in the whole system remains constant. To derive a new kinetic equation considering this facts, we proposed to replace the interaction term that accounts for every single collision event, by an interaction potential in which the colliding particles arising from the ground state are treated within the Bogoliubov approximation. This means that the creation and annihilation operators involving the ground state $a_0^\dagger$ and  $a_0$ can be substituted by $N_0$, being $N_0$ the number of particles in the condensate. Such an assumption lead us to end up with an equation, Eq. (\ref{eccin2}), that describes the dynamics below $T_c$ for which the chemical potential no longer varies, but is equal to zero. 

We have presented a numerical simulation employing equations (\ref{eccin}) and (\ref{eccin2}) to model the process of the Bose-Einstein condensation. To achieve this purpose we implemented an energetic cooling protocol. This scheme consist of maintaining fixed the number of particles during the whole cooling process while several energetic cooling stages lead the system to decrease its temperature. To lower the temperature we considered a given number of particles distributed according to the Bose-Einstein statistics, then, we removed the particles occupying the highest excited states and distribute them uniformly in the remaining lowest states. The system was allowed to evolve via the kinetic equations towards the equilibrium state. The process of particle redistribution is repeated successively to follow the evolution towards the condensation. In our numerical simulation we were able to handle with up to 40 energy quantum levels of the harmonic oscillator in 3D. To the best of our knowledge this is the highest energy level numerically computed up to now. We solved the system of coupled kinetic equations considering the ergodic assumption. The cooling sequence allowed us to follow the sequence of the growing of particles in the ground and excited states.

{\bf Acknowledgments}

\noindent
This work was partially funded by grant IN107014 DGAPA (UNAM). A.C.G. and M.M.L. acknowledges scholarship from CONACYT.


\begin{thebibliography}{}
\bibitem{Euler} To solve the coupled system we use the Euler method, selecting an appropriate temporal step to ensure the conservation of the total energy and the total number of particles.
\bibitem{Numerics} The criterion that we use to identify the situation of equilibrium at each cooling stage, is that in two consecutive time steps the value of each $n_m(\tau)$ does not change in more than $10^{-12}$.
\bibitem{N_T_mu} Such effect is a consequence of demanding that the majority of particles participate in the dynamics, that is, that they occupy states below ${\mathcal N}$=40. For $N=10^3,10^4,10^5$ and  $10^6$, the values of $T$ and $\mu$  are $(T=0.9, \mu =-0.00874)$, $(T=0.661, \mu=-0.0000874)$, $(T=0.5, \mu=-0.0000053)$ and $(T=0.5, \mu=-0.00000053)$ respectively.
\end{thebibliography}
\end{document}